\documentclass[11pt,twoside]{article}

\usepackage{graphicx}
\usepackage{asp2006}
\usepackage{epsf}
\usepackage{psfig}
\usepackage{lscape}

\begin{document}
\title{AKARI/IRC observations of heavily obscured oxygen-rich AGB and post-AGB stars}   
\author{F. Bunzel$^1$, D.A. Garc\'{i}a-Hern\'{a}ndez$^2$, D. Engels$^1$, J.V. Perea-Calder\'{o}n$^3$, and  P. Garc\'{i}a-Lario$^3$}   
\affil{$^1$ Hamburger Sternwarte, Universit\"at Hamburg, Germany \\
$^2$ Instituto de Astrof\'{i}sica de Canarias, Tenerife, Spain \\
$^3$ European Space Astronomy Centre, ESAC, ESA, Madrid, Spain \\ 
}    

\begin{abstract} 
We present AKARI/IRC observations of a sample of six extremely red
IRAS sources, of which three are variable OH/IR stars and the rest are
early post-AGB stars. The OH/IR stars show a red continuum with the
expected strong 10$\mu$m silicate absorption feature, while the
post-AGB stars show an even redder continuum accompanied with a
comparably weak silicate absorption. We modelled the spectral energy
distributions with DUSTY. While for the OH/IR stars a reasonable fit
can be obtained with almost pure silicate dust, the post-AGB stars
require a mixture of silicate and carbon-rich dust. We assume that
in the latter objects the inner dust shell is carbon-rich, while the
outer shells are still oxygen-rich.
\end{abstract}


\section{Introduction}   
During the latest phases of stellar evolution on the Asymptotic Giant
Branch (AGB), stars experience high mass loss rates of up to $10^{-4}$
$M_{\odot}$/yr and form dense optically thick circumstellar envelopes (CSE)
made up of dust and gas. The stars are pulsating as Mira variables but
with much longer periods. They are completely obscured and discovered
by OH maser or infrared surveys as OH/IR stars or "extreme" carbon
stars. At the tip of the AGB, when except of the core mass, all mass
has been lost, pulsation ceases and the remnant CSE dissipates into
the interstellar medium.  The departure from the AGB occurs, while the
stars (the remnant cores) are still completely hidden.

We observed six oxygen-rich candidates 
selected on the base of their maser properties and their IRAS infrared
colors, to study details of the AGB - post-AGB transition phase.  All
stars selected have OH or H$_2$O masers. Three stars
(IRAS\,11549$-$6225, 14104$-$5819, OH\,30.7+0.4) are pulsating
variables still on the AGB, while the others (OH\,31.0+0.0,
IRAS\,19134+2131, OH\,77.9+0.2) are post-AGB candidates.


\begin{figure}[!ht]
\begin{center}
  {\includegraphics*[width=7cm,angle=-90]{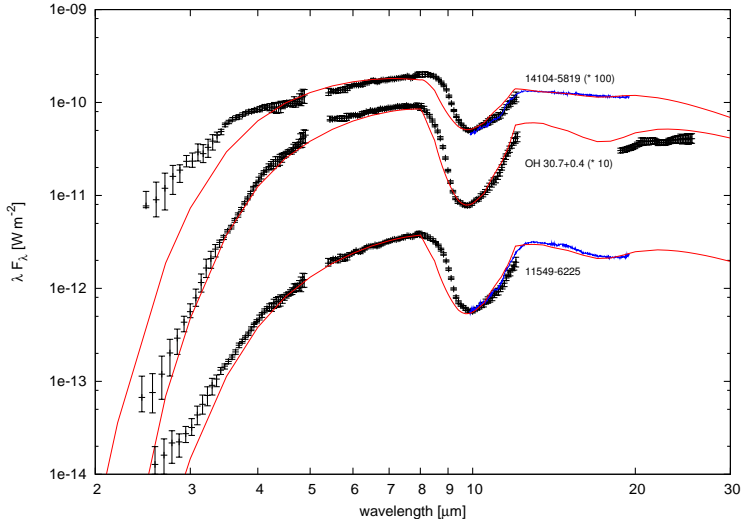}}
\end{center}
\caption{AKARI spectra
  (black) with DUSTY model fits (red) for the AGB stars.  The
  13--18$\mu$m gaps were filled with Spitzer spectra (blue). 
 The short wavelength part
of the AKARI spectrum of IRAS\,14104--5819 is contaminated by an overlapping 
spectrum.}
\label{fig:sed-agb}
\end{figure}

\section{Observations and data reduction} 
The AKARI \citep{murakami07} observations were made between
2006-10-05 and 2007-05-17.  We used the Infrared Camera (IRC) 
\citep{onaka07}
with the spectroscopic observation mode AOT\,04. We
obtained long and short exposed spectroscopic observations of
dispersion elements NP, 
SG1 and 2, 
and in one case LG2. 
The data reduction was made with the IRC Spectroscopy Toolkit Version 20080528 
\citep{ohyama07}. For some sources we encountered saturation effects,
so that the short exposures were used.
The spectra extracted by the pipeline from the calibrated images are shown
in Figures \ref{fig:sed-agb} and \ref{fig:sed-postagb}. 
We found that the
calibration was unreliable at the edges of each band and we confined
therefore the nominal wavelength ranges:  
NP (2.1--4.9$\mu$m), SG1
(5.4--8.0$\mu$m), SG2 (8.0--12.1$\mu$m) and LG2 (18.7--25.5$\mu$m).
A residual calibration error between the
spectra of dispersion elements NP and SG1 was found, probably because
of insufficient sky background removal in the crowded fields. The NP
wavelength range was therefore scaled to match the SG1 range.
Additional Spitzer spectra were taken by us under the GO program \#30258
(Garc\'{i}a-Hern\'{a}ndez et al., these proceedings). We use data from these
spectra to cover the 13--18$\mu$m range.

All six sources, AGB as well as post-AGB stars, show a clearly visible
10$\mu$m-feature of amorphous silicates, which are expected to be the
major dust component of O-rich circumstellar shells. The mid-infrared
continua of the post-AGB stars are much redder than those of the AGB
stars, indicating the presence of a more obscured shell. None of them
was found on the NP images at $\lambda < 5\mu$m.
Surprisingly, their 10$\mu$m-feature is weaker than for the AGB stars.

\section{Modelling the spectral energy distributions} 
To characterize the obtained SEDs we applied model calculations using DUSTY,
which is a 1D-radiation transport code for dusty environments
\citep{ivezic99}. DUSTY simulates a radiating source inside of a dust shell
both in spherical and planar geometry. To be able to compare the models between
the different stars and to order them by their evolutionary stage, we
fixed as much model parameters as possible. \citet{suh99} proposed
to use the optical depth as the only variable parameter, assuming that
the progress on the AGB evolutionary track depends only on the amount
of mass it has recently lost. We found that with this assumption it is
not possible to model the transition from AGB to post-AGB. Apparently
also the dust chemistry is changing during the transition.  Therefore,
as a second parameter the chemical composition of the dust had to be
changed as well. All other parameters were fixed and set to the
following values to create a 'standard model': Effective temperature
T$_{eff} = 2500$\,K, dust condensation temperature T$_{d} = 1000$\,K,
a single grain size of 0.27$\mu$m, and a density distribution obtained
from the hydrodynamic calculation option given in the code. 
The optical constants of the dust were taken from \citet{suh99} for
amorphous silicates and from \citet{suh00} for amorphous carbon dust.

\vspace{-0.3cm}
\begin{table}[!ht]
\caption{Varied model parameters for investigated AGB and post-AGB stars}
\label{tbl:agb-tab}
\begin{center}
\begin{tabular} {llllllll}
\hline
\hline
Source  & Class. &  $\tau_{10\mu m}$ &   $\tau_{10\mu m}$  &   $\tau_{10\mu m}$   &  $\eta_{sil}$ &  $\eta_{C}$ &  T$_{outer}$  \\
        &        &                   &      inner          &   outer              &    [\%]       &       [\%]  & [K]  \\
\hline
IRAS\,14104$-$5819  & AGB &   9   &  --- & --- &  90    & 10 & --- \\
IRAS\,11549$-$6225  & AGB &  16   &  --- & --- &  86    & 14 & --- \\
OH\,30.7+0.4        & AGB &  18   &  --- & --- &  97    & 03 & --- \\
OH\,77.9+0.2      & post-AGB & --- &   0.3      &   4        &    26  &     74 &   160 \\
IRAS\,19134+2131  & post-AGB & --- &   1        &   5        &    37  &     63 &   200 \\
OH\,31.0+0.0      & post-AGB & --- &    5       &   5        &    40  &     60 &   140 \\
\hline
\end{tabular}
\end{center}
\end{table}

\vspace{-0.3cm}
For the three AGB stars we created a silicate-only DUSTY model varying
the optical depth until the best fit was obtained for the strength of
the 10$\mu$m-feature and for the continuum. The remaining differences
between observed SED and model fit were reduced by adding amorphous
carbon dust and readjusting the optical depth. Figure \ref{fig:sed-agb}
shows the final fits of the three AGB stars. Table \ref{tbl:agb-tab}
(Class: AGB) gives the associated model parameters.

For the three post-AGB stars we assumed that the mass loss has
recently decreased strongly and that the inner part of the shell is
composed of amorphous carbon. A second outer shell was assumed to
contain a mixture of amorphous silicate and carbon grains. The inner
shell was modelled as a spherical shell, with an inner dust condensation
temperature of 1000\,K and a density distribution derived from
hydrodynamical calculation. The output
spectrum was used as input to the outer shell modelled in slab
geometry. In this configuration the model mimics a carbon star seen
through a screen containing silicate dust.

\begin{figure}[!ht]
\begin{center}
 {\includegraphics*[width=7cm,angle=-90]{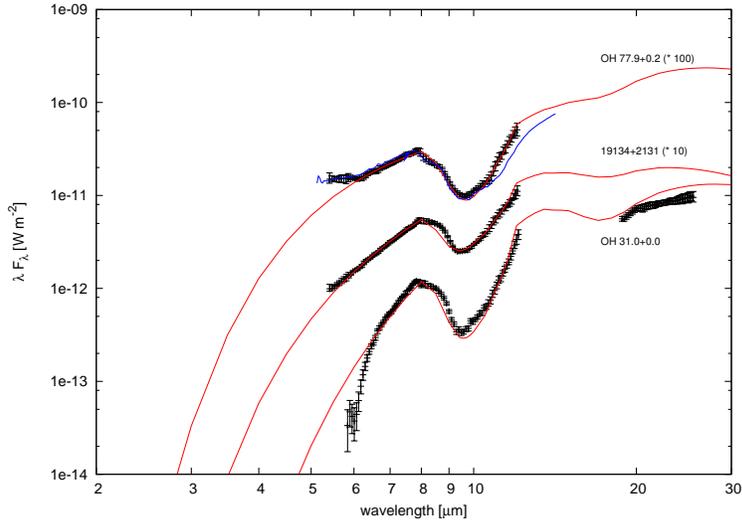}}
\end{center}
\caption{AKARI spectra
  (black) with DUSTY model fits (red) for the post-AGB stars.  The
  13--18$\mu$m gaps were filled with Spitzer spectra (blue).  The
  spectrum of OH\,77.9+0.2 is contaminated for $\lambda<6\mu$m.
}\label{fig:sed-postagb}
\end{figure}

Table \ref{tbl:agb-tab} (Class: post-AGB) gives for the best fits the
values of the model parameters varied. $\tau_{10\mu m}$ gives the
optical depths of the inner C-rich shell and the outer slab of mixed
chemistry.  T$_{outer}$ is the temperature at the inner border of
the slab.  $\eta_{sil}$ and $\eta_{C}$ are the fractions of silicate
and amorphous carbon dust. Figure \ref{fig:sed-postagb} shows the
final fits of the post-AGB stars.

\section{Discussion}
The AKARI spectra clearly show that the mid-infrared spectra of stars
departing from the AGB are much redder than those of pulsating
OH/IR stars. This is not unexpected as the post-AGB stars should
be surrounded by a remnant shell representing for a given mass the
highest mass loss rate achievable on the AGB. Using pure
silicate dust, high optical depths are needed to fit the red continua
of the post-AGB stars. Such models, however, predict a much stronger
10$\mu$m absorption feature than observed. By assuming that the red
continuum is made by carbon-rich dust and the 10$\mu$m absorption
is due to additional silicate-rich dust in the outer shell, a
reasonable fit can be found.

As all the post-AGB stars still host masers in an oxygen-rich
environment, the conversion of the inner shell chemistry cannot have
happened more than a few thousand years ago. All stars must have
experienced the conversion into a carbon star, while they were
losing the last amounts of matter of their envelopes before
exposing their cores.


\acknowledgements This research is based on observations with AKARI, a
JAXA project with the participation of ESA, and on observations with
Spitzer, a NASA's Great Observatories Program. DE acknowledges travel
support by the conference organizers.


\end{document}